\documentclass[letterpaper,english, prl, longbibliography, reprint, aps]{revtex4-1}
\usepackage[T1]{fontenc}
\usepackage[latin9]{inputenc}
\usepackage{babel}
\usepackage{amsmath}
\usepackage{amssymb}
\usepackage{graphicx}
\usepackage[unicode=true,pdfusetitle,
 bookmarks=true,bookmarksnumbered=false,bookmarksopen=false,
 breaklinks=false,pdfborder={0 0 1},backref=false,colorlinks=false]
 {hyperref}

\makeatletter

\usepackage{xcolor}

\makeatother

\begin{document}
\title{Structure-specific, mode-resolved phonon coherence and specularity
at graphene grain boundaries}
\author{Zhun-Yong Ong}
\affiliation{Institute of High Performance Computing, A{*}STAR, Singapore 138632,
Singapore}
\email{ongzy@ihpc.a-star.edu.sg}

\author{Georg Schusteritsch}
\affiliation{Department of Materials Science and Metallurgy, University of Cambridge,
27 Charles Babbage Road, Cambridge CB3 0FS, United Kingdom}
\affiliation{Advanced Institute for Materials Research, Tohoku University 2-1-1
Katahira, Aoba, Sendai, 980-8577, Japan}
\author{Chris J. Pickard}
\affiliation{Department of Materials Science and Metallurgy, University of Cambridge,
27 Charles Babbage Road, Cambridge CB3 0FS, United Kingdom}
\affiliation{Advanced Institute for Materials Research, Tohoku University 2-1-1
Katahira, Aoba, Sendai, 980-8577, Japan}
\date{\today}
\begin{abstract}
In spite of their importance for understanding phonon transport phenomena
in thin films and polycrystalline solids, the effects of boundary
roughness scattering on phonon specularity and coherence are poorly
understood because there is no general method for predicting their
dependence on phonon momentum, frequency, branch and boundary morphology.
Using the recently formulated atomistic S-matrix method, we develop
a theory of boundary roughness scattering to determine the mode-resolved
phonon coherence and specularity parameters from the scattering amplitudes.
To illustrate the theory, we apply it to phonon scattering in realistic
nonsymmetric graphene grain boundary (GB) models derived from atomic
structure predictions. The method is validated by comparing its predictions
with frequency-resolved results from lattice dynamics-based calculations.
We prove that incoherent scattering is almost perfectly diffusive.
We show that phonon scattering at the graphene GB is not diffuse although
coherence and specularity are significantly reduced for long-wavelength
flexural acoustic phonons. Our approach can be generalized to other
atomistic boundary models.
\end{abstract}
\maketitle

\newcommand{\SuppInfo}[1]{\textcolor{blue}{\emph{SI Appendix} #1}}

Phonon mean free path (MFP) engineering through boundary roughness
scattering is a widely used approach to manipulating phonon transport
in low-dimensional materials (e.g. silicon nanowires~\citep{AHochbaum:Nature08_Thermoelectric,JLim:NL12_Quantifying})
for thermoelectric and thermal management applications~\citep{DLi:NMTE15_Phonon,CMonachon:ARMR16_Thermal}
as well as for investigations into fundamental phonon phenomena such
as phonon hydrodynamics~\citep{RAGuyer:PR66_Solution} in layered
crystals~\citep{ZDing:NL18_Phonon} and ballistic phonons in graphene~\citep{MBae:NatCommun13_Ballistic}.
In nanostructures, the reduced thermal conductivity is also attributed
to boundary roughness scattering~\citep{GChen:JHT97_Size,GChen:Book05_Nanoscale}.
Nonetheless, in spite of its importance for phonon transport, a rigorous
quantitative description of how phonons undergo momentum and phase
relaxation from boundary roughness scattering still eludes us~\citep{GChen:Book05_Nanoscale,DLi:NMTE15_Phonon},
posing an obstacle to the systematic use of structural modification
to control the phonon MFP, while a direct characterization of the
specularity is very difficult with current experimental techniques~\citep{NRavichandran:PRX18_Spectrally}.
Although there have been studies using phonon wavepackets to probe
boundary scattering~\citep{PKSchelling:APL02_Phonon,PKSchelling:JAP04_Kapitza,LNMaurer:PRB16_Rayleigh,CShao:JAP17_Probing,CShao:PRB18_Understanding},
their use is limited by the considerable difficulty of deriving mode-resolved
reciprocal-space information from real-space data in addition to the
substantial computational costs.

\begin{figure}
\begin{centering}
\includegraphics[scale=0.27]{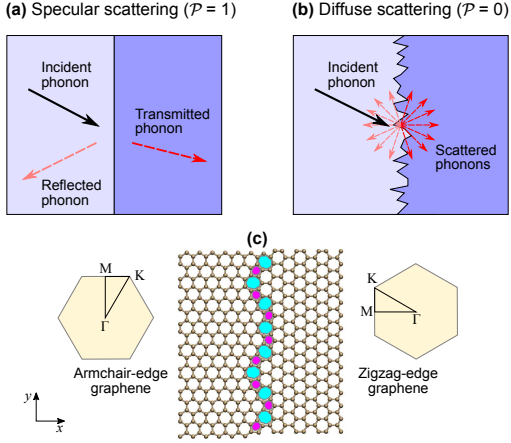}
\par\end{centering}
\caption{Depiction of \textbf{(a)} perfectly specular versus \textbf{(b)} perfectly
diffuse scattering at a boundary, and \textbf{(c)} the graphene GB
between armchair- and zigzag-edge graphene. The shape and orientation
of their respective Brillouin zones are also shown.}
\label{fig:ScatteringCartoon}
\end{figure}

A major challenge to understanding this mechanism is our inability
to predict accurately for a given boundary model the probability of
the incident phonon undergoing specular scattering, characterized
by the specularity parameter $\mathcal{P}$ which plays an important
role in many boundary scattering models~\citep{ZAksamija:PRB10_Anisotropy,JLim:NL12_Quantifying,AMajee:PRB16_Length}
and should vary with phonon frequency, momentum and polarization/branch.
In perfectly specular scattering ($\mathcal{P}=1$) as shown in Fig.~\ref{fig:ScatteringCartoon}(a),
the incident bulk phonon is scattered coherently by a smooth boundary
into well-defined trajectories while in perfectly diffuse scattering
($\mathcal{P}=0$) or the so-called Casimir limit as shown in Fig.~\ref{fig:ScatteringCartoon}(b),
the incoming phonon energy is redistributed uniformly over the entire
spectrum of outgoing phonon channels, resulting in maximum momentum
loss in the direction parallel to the boundary~\citep{GChen:Book05_Nanoscale}.
Another challenge lies in predicting the effect of boundary roughness
on coherent and incoherent scattering, an unresolved issue in phonon
transport in superlattices where the role of phonon interference in
thermal conductivity is still debated~\citep{BYang:PRB03_Partially,MNLuckyanova:Science12_Coherent,JRavichandran:NatMater14_Crossover,YWang:PRB14_Decomposition}.

In order to address these challenges, we develop in this paper a theory
of boundary roughness scattering, based on the recently formulated
atomistic $S$-matrix method~\citep{ZYOng:PRB18_Atomistic}, to determine
the \emph{mode-resolved} phonon coherence and specularity parameters
for boundary models. Unlike existing approaches~\citep{AMaznev:PRB15_Boundary,FShi:PRB17_Diffusely},
our method is fully atomistic, not restricted to long-wavelength modes,
and distinguishes coherent and incoherent scattering~\citep{MLax:RMP51_Multiple,AIshimaru:Book78_Wave_Chap4,JAOgilvy:RPR87_Wave}
by treating boundary roughness in a statistical manner analogous to
the theory of multiple scattering in disordered systems~\citep{LFoldy:PR45_Multiple,MLax:RMP51_Multiple,VTwersky:JASA57_Scattering,VTwersky:JMP62_Scattering}
and conceptually similar to the approach in Ref.~\citep{DKechrakos:JPCM90_Phonon}.
We apply this theory to phonon scattering at the grain boundary (GB)
between armchair- and zigzag-terminated graphene like in Fig.~\ref{fig:ScatteringCartoon}(c),
using realistic nonsymmetric low-energy GB models derived from \emph{ab
initio}-based structure predictions~\citep{GSchusteritsch:PRB14_Predicting}.
We validate our method by comparing its predictions with the less
precise Zhao-Frend method~\citep{HZhao:JAP09_Phonon} and analyze
how the coherence and specularity parameters vary with phonon frequency,
momentum and polarization/branch for the graphene GB.

\section{Theory and model}

\subsection{Grain boundary model and $S$ matrix}

\begin{figure}
\begin{centering}
\includegraphics[scale=0.27]{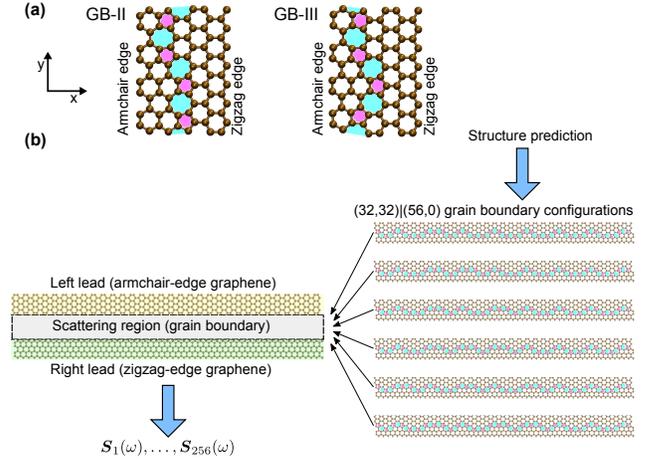}
\par\end{centering}
\caption{\textbf{(a)} Atomistic structure of the (4,4)|(7,0) GB-II and GB-III
interfaces. \textbf{(b)} Schematic of atomistic $S$-matrix calculation
with the scattering region comprising the (32,32)|(56,0) grain boundary
(GB). We generate an ensemble of 256 GB configurations derived from
structure predictions. Each GB configuration is inserted into the
scattering region between the left and right leads and its corresponding
$S$ matrix is computed using Ref.~\citep{ZYOng:PRB18_Atomistic}.}
\label{fig:SimulationScheme}
\end{figure}

To treat phonon scattering by the rough (32,32)|(56,0) graphene GB
statistically, we need to generate the various possible GB configurations
and their interatomic force constant (IFC) matrices. Each (32,32)|(56,0)
graphene GB configuration, which consists of an undulating line of
pentagon-heptagon defect pairs like in Fig.~\ref{fig:ScatteringCartoon}(c),
is constructed from an 8-unit random sequence of the two lowest-energy
(4,4)|(7,0) graphene GB configurations (GB-II and GB-III in Fig.~\ref{fig:SimulationScheme}(a))
in Ref.~\citep{GSchusteritsch:PRB14_Predicting}, with open-system
and periodic boundary conditions in the $x$ and $y$ direction, respectively,
to yield $2^{8}=256$ unique GB configurations. Given the large size
of the GB models, we use the program GULP~\citep{JGale:MolSim03_gulp}
and the empirical Tersoff potential~\citep{JTersoff:PRL88_Empirical},
with parameters from Ref.~\citep{LLindsay:PRB10_Optimized}, to model
the C-C interatomic forces instead of more expensive \emph{ab initio}
methods and to compute the IFC matrices needed for the atomistic $S$-matrix
calculations as described in Ref.~\citep{ZYOng:JAP18_Tutorial,ZYOng:PRB18_Atomistic},
with details of the GB structure generation and optimization given
in Sec.~S1 of the Supplemental Material~\citep{ZYOng:SuppMater19_Structure}.
The scheme of our calculations is shown in Fig.~\ref{fig:SimulationScheme}(b).

Using our code which implements the atomistic $S$-matrix method~\citep{ZYOng:PRB18_Atomistic},
we compute at each frequency $\omega=n\omega_{0}$, where $n=1,\ldots,25$
and $\omega_{0}=10^{13}$ rad/s, the unitary $N(\omega)\times N(\omega)$
matrix $\boldsymbol{S}(\omega)$ which describes the mapping of the
$N(\omega)$ incoming bulk phonon modes to the $N(\omega)$ outgoing
bulk phonon modes on both sides of the boundary, for each GB configuration.
Details of the $S$-matrix calculations are given in Sec.~S2 of the
Supplemental Material~~\citep{ZYOng:SuppMater19_Structure}. In
the general scattering picture~\citep{ZYOng:JAP18_Tutorial,ZYOng:PRB18_Atomistic},
$\boldsymbol{S}(\omega)$, which relates the incoming phonon state
$\boldsymbol{\Phi}_{\text{in}}$ to the outgoing phonon state $\boldsymbol{\Psi}_{\text{out}}$
via the relation $\boldsymbol{\Psi}_{\text{out}}=\boldsymbol{S}(\omega)\boldsymbol{\Phi}_{\text{in}}$,
encodes the amplitude and phase changes. Numerically, $\boldsymbol{\Phi}_{\text{in}}$
and $\boldsymbol{\Psi}_{\text{out}}$, which represent a superposition
of $N(\omega)$ bulk phonon modes, are column vectors with the $m$-th
element of $\boldsymbol{\Phi}_{\text{in}}$ ($\boldsymbol{\Psi}_{\text{out}}$)
equal to the complex flux amplitude of the $m$-th incoming (outgoing)
phonon channel and represented by $[\boldsymbol{\Phi}_{\text{in}}]_{m}=\boldsymbol{\Phi}(\boldsymbol{k}_{m})$
and $[\boldsymbol{\Psi}_{\text{out}}]_{m}=\boldsymbol{\Psi}(\boldsymbol{k}_{m})$
for $m=1,\ldots,N(\omega)$ with the momentum $\boldsymbol{k}_{m}$
and branch $\nu_{m}$ associated with the $m$-th phonon channel.
We can thus interpret $|\boldsymbol{\Phi}(\boldsymbol{k}^{\prime})|^{2}$
and $|\boldsymbol{\Psi}(\boldsymbol{k})|^{2}$ as the intensity of
the incoming $\boldsymbol{k}^{\prime}$ and the outgoing $\boldsymbol{k}$
phonon flux, respectively. Hence, the matrix element $[\boldsymbol{S}(\omega)]_{mn}=S(\boldsymbol{k}_{m},\boldsymbol{k}_{n}^{\prime})$
is equal to the scattering amplitude from the $n$-th incoming to
the $m$-th outgoing phonon channel, i.e.,  
\begin{align}
\begin{pmatrix}\boldsymbol{\Psi}(\boldsymbol{k}_{1})\\
\vdots\\
\boldsymbol{\Psi}(\boldsymbol{k}_{N})
\end{pmatrix}=\left(\begin{array}{ccc}
S(\boldsymbol{k}_{1},\boldsymbol{k}_{1}^{\prime}) & \ldots & S(\boldsymbol{k}_{1},\boldsymbol{k}_{N}^{\prime})\\
\vdots & \ddots & \vdots\\
S(\boldsymbol{k}_{N},\boldsymbol{k}_{1}^{\prime}) & \ldots & S(\boldsymbol{k}_{N},\boldsymbol{k}_{N}^{\prime})
\end{array}\right)\begin{pmatrix}\boldsymbol{\Phi}(\boldsymbol{k}_{1}^{\prime})\\
\vdots\\
\boldsymbol{\Phi}(\boldsymbol{k}_{N}^{\prime})
\end{pmatrix}\label{eq:SMatrix}
\end{align}
where $\{\boldsymbol{k}_{1},\ldots,\boldsymbol{k}_{N(\omega)}\}$
and $\{\boldsymbol{k}_{1}^{\prime},\ldots,\boldsymbol{k}_{N(\omega)}^{\prime}\}$
denote the momenta of the outgoing and incoming modes, respectively.

The evaluation of Eq.~(\ref{eq:EnsembleSMatrices}) requires a configurational
ensemble of \emph{$S$} matrices computed using the method described
in Ref~\citep{ZYOng:PRB18_Atomistic}, with each matrix describing
a boundary configuration. For simplicity, we choose the (32,32)|(56,0)
graphene GB as our boundary model which we construct from the two
lowest-energy (4,4)|(7,0) GB configurations (GB-II and GB-III in Fig.~\ref{fig:SimulationScheme}(a))
in Ref.~\citep{GSchusteritsch:PRB14_Predicting} found using the
\emph{ab initio} random structure searching method~\citep{CPickard:PRL06_High}.
Each (32,32)|(56,0) GB configuration comprises eight (4,4)|(7,0) GB's,
a permutation of GB-II's and GB-III's, forming a continuous line of
pentagon-heptagon defect pairs. This construction method yields $2^{8}=256$
unique GB configurations. We set the direction of the phonon flux
and the GB to be parallel to the $x$- and $y$-axis, respectively
and impose periodic boundary conditions in the $y$-direction. Given
the large size of the GB models, we use the empirical Tersoff potential~\citep{JTersoff:PRL88_Empirical},
with parameters from Ref.~\citep{LLindsay:PRB10_Optimized}, to model
the C-C interatomic forces instead of more expensive \emph{ab initio}
methods. The program GULP~\citep{JGale:MolSim03_gulp} is used to
optimize each GB configuration and to generate its force-constant
matrices $\boldsymbol{H}_{\text{CL}}$, $\boldsymbol{H}_{\text{C}}$
and $\boldsymbol{H}_{\text{CR}}$ needed for the $S$-matrix calculations.
We also compute the force-constant matrices $\boldsymbol{H}_{\text{L}}^{00}$
and $\boldsymbol{H}_{\text{L}}^{01}$ ($\boldsymbol{H}_{\text{R}}^{00}$
and $\boldsymbol{H}_{\text{R}}^{01}$) describing the armchair-edge
(zigzag-edge) graphene in the left (right) lead. At each frequency
$\omega=n\omega_{0}$ ($n=1,\ldots,25$ and $\omega_{0}=10^{13}$
rad/s), we compute an $N(\omega)\times N(\omega)$ matrix $\boldsymbol{S}_{\alpha}(\omega)$
for the $\alpha$-th GB configuration ($\alpha=1,\ldots,256$).

\subsection{$S$-matrix theory of boundary roughness scattering}

For a nonideal boundary that consists of a deterministic part corresponding
to the smooth boundary and a stochastic part describing the boundary
roughness, $\boldsymbol{\Psi}_{\text{out}}$ can be partitioned into
its deterministic and stochastic components in a manner akin to the
treatment of randomly scattered wave fields~\citep{MLax:RMP51_Multiple,AIshimaru:Book78_Wave_Chap4,JAOgilvy:RPR87_Wave},
i.e., 
\begin{equation}
[\boldsymbol{\Psi}_{\text{out}}]_{m}=\langle[\boldsymbol{\Psi}_{\text{out}}]_{m}\rangle+[\delta\boldsymbol{\Psi}_{\text{out}}]_{m}\label{eq:SmoothRoughPhononStates}
\end{equation}
where $\langle[\boldsymbol{\Psi}_{\text{out}}]_{m}\rangle$ and $[\delta\boldsymbol{\Psi}_{\text{out}}]_{m}$
are its deterministic and stochastic components, respectively, and
$\langle\ldots\rangle$ represents the configurational average~\citep{PCWaterman:JMP61_Multiple}
assuming that every configuration is equally probable. Similarly,
the deterministic and stochastic components of $\boldsymbol{S}(\omega)$
are defined via the expression $[\boldsymbol{S}(\omega)]_{mn}=\langle[\boldsymbol{S}(\omega)]_{mn}\rangle+[\delta\boldsymbol{S}(\omega)]_{mn}$
where $\langle[\boldsymbol{\Psi}_{\text{out}}]_{m}\rangle=\sum_{n=1}^{N}\langle[\boldsymbol{S}(\omega)]_{mn}\rangle[\boldsymbol{\Phi}_{\text{in}}]_{n}$
and $[\delta\boldsymbol{\Psi}_{\text{out}}]_{m}=\sum_{n=1}^{N}[\delta\boldsymbol{S}(\omega)]_{mn}[\boldsymbol{\Phi}_{\text{in}}]_{n}$.
For any given $[\boldsymbol{\Phi}_{\text{in}}]_{n}$, the deterministic
component $\langle[\boldsymbol{\Psi}_{\text{out}}]_{m}\rangle$ and
hence $\langle[\boldsymbol{S}(\omega)]_{mn}\rangle$ preserve the
coherent amplitude and phase information from direct averaging.

It follows from Eq.~(\ref{eq:SmoothRoughPhononStates}) that $\langle[\delta\boldsymbol{\Psi}_{\text{out}}]_{m}\rangle=0$,
i.e., the \emph{amplitude} fluctuations of the outgoing phonon state
average to zero, and thus $\langle[\delta\boldsymbol{S}(\omega)]_{mn}\rangle=0$.
However, the configurational average of $|[\boldsymbol{\Psi}_{\text{out}}]_{m}|^{2}$,
the \emph{probability} of the phonon being scattered to the $m$-th
outgoing phonon channel, is $\langle|[\boldsymbol{\Psi}_{\text{out}}]_{m}|^{2}\rangle=|\langle[\boldsymbol{\Psi}_{\text{out}}]_{m}\rangle|^{2}+\langle|[\delta\boldsymbol{\Psi}_{\text{out}}]_{m}|^{2}\rangle$,
implying that the transition \emph{probability} fluctuations associated
with boundary roughness are not necessarily zero since $\langle|[\boldsymbol{\Psi}_{\text{out}}]_{m}|^{2}\rangle\geq|\langle[\boldsymbol{\Psi}_{\text{out}}]_{m}\rangle|^{2}$.
Hence, the configurational average of the transition probability is
given by $\langle|[\boldsymbol{S}(\omega)]_{mn}|^{2}\rangle=|\langle[\boldsymbol{S}(\omega)]_{mn}\rangle|^{2}+\langle|[\delta\boldsymbol{S}(\omega)]_{mn}|^{2}\rangle$,
which we rewrite as $[\boldsymbol{W}_{\text{total}}(\omega)]_{mn}=[\boldsymbol{W}_{\text{coh}}(\omega)]_{mn}+[\boldsymbol{W}_{\text{incoh}}(\omega)]_{mn}$
where $\boldsymbol{W}_{\text{total}}$ , $\boldsymbol{W}_{\text{coh}}$
and $\boldsymbol{W}_{\text{incoh}}$ are the total, coherent and incoherent
transition probability matrices, respectively, with their matrix elements
given by  
\begin{subequations}
\begin{align}
[\boldsymbol{W}_{\text{total}}(\omega)]_{mn} & =\langle|[\boldsymbol{S}(\omega)]_{mn}|^{2}\rangle\label{eq:EnsembleRMSSMatrix}\\{}
[\boldsymbol{W}_{\text{coh}}(\omega)]_{mn} & =|\langle[\boldsymbol{S}(\omega)]_{mn}\rangle|^{2}\label{eq:EnsembleCoherentSMatrix}\\{}
[\boldsymbol{W}_{\text{incoh}}(\omega)]_{mn} & =\langle|[\boldsymbol{S}(\omega)]_{mn}|^{2}\rangle-|\langle[\boldsymbol{S}(\omega)]_{mn}\rangle|^{2}\ .\label{eq:EnsembleIncoherentSMatrix}
\end{align}
\label{eq:EnsembleSMatrices}
\end{subequations}
 $[\boldsymbol{W}_{\text{total}}(\omega)]_{mn}$
represents the \emph{total} transition probability between the $n$-th
incoming and the $m$-th outgoing channel while $[\boldsymbol{W}_{\text{coh}}(\omega)]_{mn}$
and $[\boldsymbol{W}_{\text{incoh}}(\omega)]_{mn}$ correspond to
its coherent and incoherent components.

\subsection{Definition of mode-resolved phonon coherence and specularity}

To characterize the coherence and specularity of the $n$-th incoming
phonon channel, we use the transition probabilities from Eq.~(\ref{eq:EnsembleSMatrices})
to define the phonon coherence $\mathcal{C}_{n}$ 
\begin{equation}
\mathcal{C}_{n}(\omega)=\sum_{m=1}^{N(\omega)}[\boldsymbol{W}_{\text{coh}}(\omega)]_{mn}\ ,\label{eq:PhononCoherence}
\end{equation}
the sum of the coherent transition probabilities, as its probability
of being coherently scattered. Equation~(\ref{eq:PhononCoherence})
satisfies $0<\mathcal{C}_{n}\leq1$ and can be interpreted as the
proportion of the incoming phonon flux redistributed to the outgoing
phonon channels after coherent scattering We recall that the specularity
parameter is the probability that the incident phonon is scattered
into the outgoing phonon channels associated with specular scattering
by an ideal boundary. Given that the structural randomness of the
rough boundary results in both coherent and incoherent scattering,
we can characterize the specularity of each type of scattering independently.
To estimate the specularity parameter associated with each type of
out-scattering from the $n$-th incoming phonon channel at frequency
$\omega$, we propose a statistical characterization of the `spread'
in the transition probabilities, given by  
\begin{subequations}
\begin{align}
P_{n}^{\text{total}}(\omega) & =\frac{\sqrt{\sum_{m=1}^{N(\omega)}|[\boldsymbol{W}_{\text{total}}(\omega)]_{mn}|^{2}}}{\sum_{m=1}^{N(\omega)}[\boldsymbol{W}_{\text{total}}(\omega)]_{mn}}\label{eq:TotalSpecularityEstimate}\\
P_{n}^{\text{coh}}(\omega) & =\frac{\sqrt{\sum_{m=1}^{N(\omega)}|[\boldsymbol{W}_{\text{coh}}(\omega)]_{mn}|^{2}}}{\sum_{m=1}^{N(\omega)}[\boldsymbol{W}_{\text{coh}}(\omega)]_{mn}}\label{eq:CoherentSpecularityEstimate}\\
P_{n}^{\text{incoh}}(\omega) & =\frac{\sqrt{\sum_{m=1}^{N(\omega)}|[\boldsymbol{W}_{\text{incoh}}(\omega)]_{mn}|^{2}}}{\sum_{m=1}^{N(\omega)}[\boldsymbol{W}_{\text{incoh}}(\omega)]_{mn}}\label{eq:IncoherentSpecularityEstimate}
\end{align}
\label{eq:SpecularityEstimates}
\end{subequations}
 where $P_{n}^{\text{coh}}$, $P_{n}^{\text{incoh}}$
and $P_{n}^{\text{total}}$ represent the coherent, incoherent and
total specularity, respectively. Equation~(\ref{eq:SpecularityEstimates})
corresponds to the normalized second moment of the transition probabilities,
satisfying $0<P_{n}^{\text{total}},P_{n}^{\text{coh}},P_{n}^{\text{incoh}}\leq1$,
and is related to the inverse participation ratio used to characterize
disordered eigenstates in Anderson localization theory~\citep{JTEdwards:JPCSSP72_Numerical}.
The numerator in Eq.~(\ref{eq:SpecularityEstimates}) counts the
effective number of outgoing channels over which the scattered energy
is distributed and measures how evenly it is spread across the outgoing
(transmitted and reflected) channels in different branches. The specularity
parameters are related to the coherence from Eq.~(\ref{eq:PhononCoherence})
through the compact expression   
\begin{equation}
(P_{n}^{\text{total}})^{2}=\mathcal{C}_{n}^{2}(P_{n}^{\text{coh}})^{2}+(1-\mathcal{C}_{n})^{2}(P_{n}^{\text{incoh}})^{2}\ .\label{eq:CoherenceSpecularityRelation}
\end{equation}

We motivate Eq.~(\ref{eq:SpecularityEstimates}) from the advantages
and consistency of its asymptotic ($N\rightarrow\infty$) behavior
with expected $\mathcal{P}$ values under well-defined conditions~\citep{GChen:Book05_Nanoscale}.
In the Casimir ($\mathcal{P}=0$) limit where the incoming phonon
energy is diffused uniformly over all $N$ outgoing phonon channels,
we have $P_{n}^{\text{total}}=N^{-1/2}$ so that $\lim_{N\rightarrow\infty}P_{n}^{\text{total}}=0$.
For perfectly specular reflection ($\mathcal{P}=1$), there is only
one outgoing phonon channel with a transition probability of unity
(i.e. $[\boldsymbol{W}_{\text{total}}(\omega)]_{mn}=1$ for some $m$)
and $P_{n}^{\text{total}}=1$ as expected. For partially specular
scattering ($\mathcal{P}=p$) where there is one dominant outgoing
phonon channel with transition probability $p$ and the transition
probability to each remaining channel is $\frac{1-p}{N-1}$, we obtain
$\lim_{N\rightarrow\infty}P_{n}^{\text{total}}=p$.

\section{Results and discussion}

\begin{figure}
\begin{centering}
\includegraphics[scale=0.5]{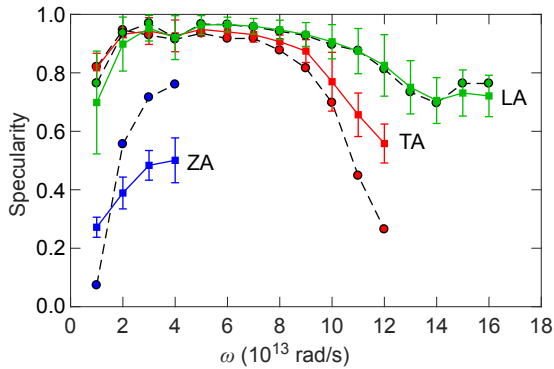}
\par\end{centering}
\caption{Comparison of the Zhao-Freund specularity parameters $p_{\alpha,\text{L}}$
(dashed lines) from Eq.~(\ref{eq:ZhaoFreundSpecularity}) with the
branch-averaged specularity parameters $\overline{P}_{\alpha,\text{L}}$
(solid lines) from Eq.~(\ref{eq:OngBranchAveragedSpecularity}) for
$\alpha=$ LA (green symbols), TA (red symbols) and ZA (blue symbols)
phonons in armchair-edge graphene.}
\label{fig:ZhaoFreundSpecularityComparison}
\end{figure}

\subsection{Comparison with Zhao-Freund specularity parameter}

In addition to its consistency under well-defined conditions, we also
validate Eq.~(\ref{eq:SpecularityEstimates}) by comparing its predictions
to the lattice dynamics-based approach from Ref.~\citep{HZhao:JAP09_Phonon}
in which Zhao and Freund define a frequency-dependent specularity
parameter $p(\omega)$, which lacks modal resolution and we may consider
as the specularity parameter averaged over all the modes in all phonon
branches at the frequency $\omega$, based on the relative value of
the actual phonon transmission to the transmission functions predicted
from the acoustic mismatch model (AMM) and diffuse mismatch model
(DMM). As we can resolve the phonon branch, we generalize the Zhao-Freund
estimate to define the more precise frequency- and \emph{branch}-dependent
specularity parameter~\citep{HZhao:JAP09_Phonon} for the left-lead
$\alpha$-branch phonons as  
\begin{equation}
p_{\alpha,\text{L}}(\omega)=\frac{\Xi_{\alpha,\text{L}}(\omega)-\Xi_{\alpha,\text{L}}^{(\text{DMM})}(\omega)}{\Xi_{\alpha,\text{L}}^{(\text{AMM})}(\omega)-\Xi_{\alpha,\text{L}}^{(\text{DMM})}(\omega)}\ \label{eq:ZhaoFreundSpecularity}
\end{equation}
 where $\alpha=$ LA (longitudinal acoustic), TA
(transverse acoustic), ZA (flexural acoustic), LO (longitudinal optical),
TO (transverse optical) or ZO (flexural optical), and $\Xi_{\alpha,\text{L}}$,
$\Xi_{\alpha,\text{L}}^{(\text{AMM})}$ and $\Xi_{\alpha,\text{L}}^{(\text{DMM})}$
are the transmission functions calculated with the atomistic $S$-matrix,
AMM and DMM method, respectively, as described in Sec\@.~S3 of the
Supplemental Material~\citep{ZYOng:SuppMater19_Structure}. We also
define the analogous branch-averaged, frequency-depedent total specularity
parameter  
\begin{equation}
\overline{P}_{\alpha,\text{L}}(\omega)=\frac{\sum_{n=1}^{N(\omega)}P_{n}^{\text{total}}(\omega)\Theta(v_{x,n}^{\prime})\delta_{\nu_{n}^{\prime},\alpha}}{\sum_{n=1}^{N(\omega)}\Theta(v_{x,n}^{\prime})\delta_{\nu_{n}^{\prime},\alpha}}\ ,\label{eq:OngBranchAveragedSpecularity}
\end{equation}
 by averaging $P_{n}^{\text{total}}$ from Eq.~(\ref{eq:SpecularityEstimates})
over all the incoming left-lead $\alpha$-branch phonon channels.
The comparison between Eqs.~(\ref{eq:ZhaoFreundSpecularity}) and
(\ref{eq:OngBranchAveragedSpecularity}) is made over the frequency
range in which we have long-wavelength phonons with momentum $\boldsymbol{k}$
satisfying $|\boldsymbol{k}|<k_{\text{cutoff}}$ where the cutoff
momentum $k_{\text{cutoff}}$ is set as half of the distance between
the $\Gamma$ and $K$-point in the first Brillouin zone (BZ).

We observe excellent agreement between $\overline{P}_{\text{LA},\text{L}}$
and $p_{\text{LA},\text{L}}$ over the entire frequency range in Fig.~\ref{fig:ZhaoFreundSpecularityComparison}.
The agreement between $\overline{P}_{\text{TA},\text{L}}$ and $p_{\text{TA},\text{L}}$
is also remarkably good although the two quantities diverge at higher
frequencies, possibly because of the deviation of the TA phonon frequencies
from the linear dispersion implicitly assumed in $\Xi_{\alpha,\text{L}}^{(\text{AMM})}$
in Eq.~(\ref{eq:ZhaoFreundSpecularity}) for estimating $p_{\text{TA},\text{L}}$.
The sensitivity of the agreement between Eqs.~(\ref{eq:ZhaoFreundSpecularity})
and (\ref{eq:OngBranchAveragedSpecularity}) to the phonon dispersion
linearity is also reflected in the poor agreement between $\overline{P}_{\text{ZA},\text{L}}$
and $p_{\text{ZA},\text{L}}$ for ZA phonons, which have a \emph{quadratic}
phonon dispersion in the long-wavelength limit in graphene~\citep{LLindsay:PRB10_Flexural},
although the general trend of the ZA phonon specularity increasing
with frequency is captured. The close agreement between Eqs.~(\ref{eq:ZhaoFreundSpecularity})
and (\ref{eq:OngBranchAveragedSpecularity}) for long-wavelength LA
and TA phonons supports our approach for estimating the specularity
parameters in Eq.~(\ref{eq:SpecularityEstimates}).

\begin{figure*}
\begin{centering}
\includegraphics[width=17.8cm]{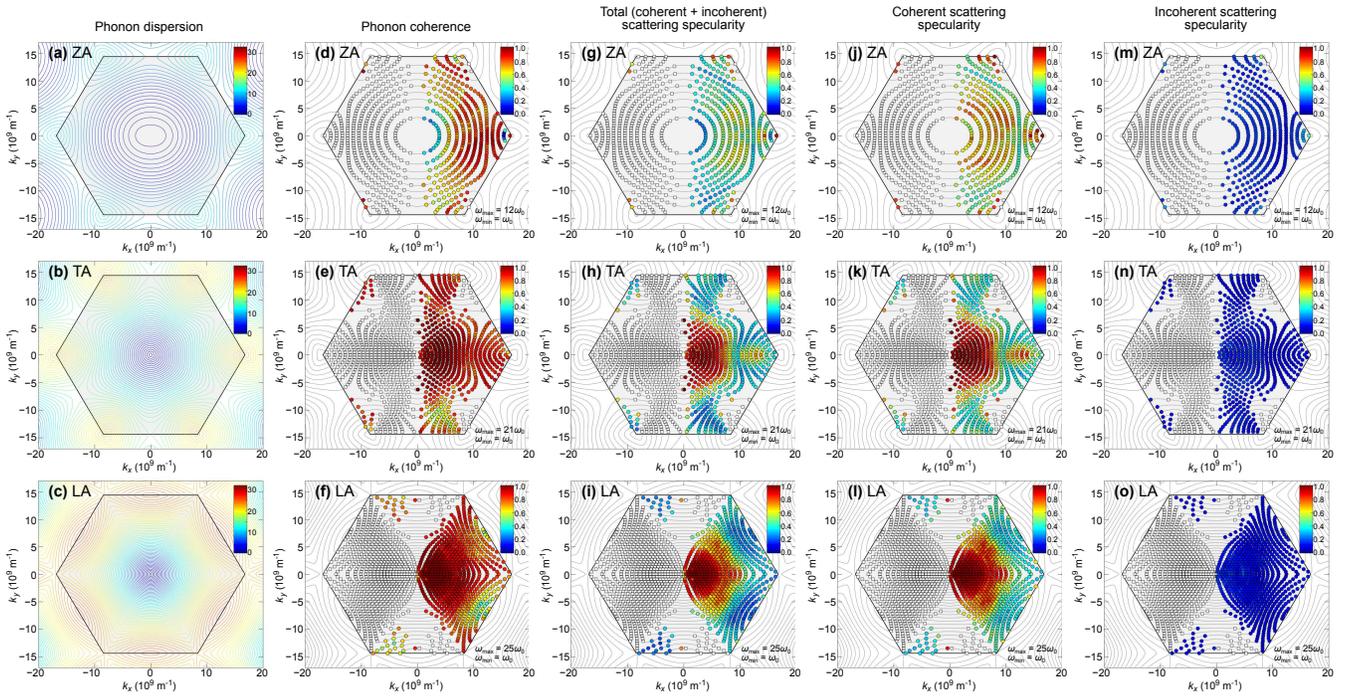}
\par\end{centering}
\caption{\textbf{(a-c)} Phonon dispersion, \textbf{(d-f)} coherence ($\mathcal{C}_{n}$)
and the estimated \textbf{(g-i)} total, \textbf{(j-l)} coherent and
\textbf{(m-o)} incoherent mode-resolved specularity parameters ($P_{n}^{\text{total}}$,
$P_{n}^{\text{coh}}$ and $P_{n}^{\text{incoh}}$) for the ZA, TA
and LA phonons in armchair-edge graphene impinging on the grain boundary.
The modes in the incoming phonon flux are filled circles colored according
to their numerical value while the modes in the outgoing flux are
hollow squares. The frequency range is $\omega=\omega_{0}$ to $25\omega_{0}$
where $\omega_{0}=10^{13}$ rad/s, with the maximum frequency ($\omega_{\text{max}}$)
for the ZA, TA and LA phonons equal $12\omega_{0}$, $21\omega_{0}$
and $25\omega_{0}$, respectively. The isofrequency contours are indicated
in intervals of $\Delta\omega=\omega_{0}$ in \textbf{(d-o)} using
solid gray lines. The phonon dispersions in \textbf{(a-c)} are indicated
with color contours in intervals of $\Delta\omega=\omega_{0}/2$.}
\label{fig:AllSpecularityParameters}
\end{figure*}

\subsection{Specularity and coherence of graphene phonons}

In Fig.~\ref{fig:AllSpecularityParameters}, we analyze the reciprocal-space
distribution of the phonon coherence ($\mathcal{C}_{n}$) and the
total, coherent and incoherent specularity parameters ($P_{n}^{\text{total}}$,
$P_{n}^{\text{coh}}$ and $P_{n}^{\text{incoh}}$) for the ZA, TA
and LA phonon modes over the entire first BZ in armchair-edge graphene,
computed from Eqs.~(\ref{eq:SpecularityEstimates}) and (\ref{eq:PhononCoherence})
over the frequency range of $\omega=\omega_{0}$ to $25\omega_{0}$
rad/s in intervals of $\omega_{0}=10^{13}$ rad/s, using the method
described in Ref.~\citep{ZYOng:PRB18_Atomistic}. The mode-resolved
data over the \emph{entire} BZ is obtained by plotting the mode-resolved
data at each frequency and then sweeping over the aforementioned frequency
range. The corresponding results for zigzag-edge graphene are omitted
here but given in Sec\@.~S4 of the Supplemental Material~\citep{ZYOng:SuppMater19_Structure}
. The convergence of $\mathcal{C}_{n}$ and $P_{n}^{\text{total}}$
with respect to GB width is also discussed in Sec. S5 of the Supplemental
Material~\citep{ZYOng:SuppMater19_Structure}.

In Fig.~\ref{fig:AllSpecularityParameters}(d), we observe that $\mathcal{C}_{n}$
for ZA phonons increases as $\boldsymbol{k}_{n}$ decreases, suggesting
that long-wavelength ZA phonons are more sensitive to GB roughness,
against conventional expectations that boundary roughness scatters
short-wavelength phonons more strongly~\citep{GChen:Book05_Nanoscale}.
In contrast, Figs.~\ref{fig:AllSpecularityParameters}(e) and (f)
show that $\mathcal{C}_{n}$ for LA and TA phonons decreases as $\boldsymbol{k}_{n}$
increases, indicating that long-wavelength LA and TA phonons are less
incoherently scattered. The trend in Fig.~\ref{fig:AllSpecularityParameters}(d)
is consistent with the $P_{n}^{\text{total}}$ distribution in Figs.~\ref{fig:AllSpecularityParameters}(g)
to (i), which show $P_{n}^{\text{total}}$ decreasing for LA and TA
phonons but increasing for ZA phonons as $\boldsymbol{k}_{n}$ increases.
We speculate that this is related to the significantly higher point-defect
scattering rates of ZA phonons in graphene~\citep{CAPolanco:PRB18_AbInitio}.
The greater GB scattering of ZA phonons implies that in suspended
polycrystalline graphene, the in-plane LA and TA phonons play a more
significant role in heat conduction than the out-of-plane ZA phonons
which are said to dominate thermal transport in pristine graphene~\citep{LLindsay:PRB10_Flexural}.
It has also been proposed by Soffer~\citep{SBSoffer:JAP67_statistical,CShao:JAP17_Probing}
that the specularity parameter should vary \emph{anisotropically}
as $P=\exp[-(2\eta k_{x})^{2}]$, where $\eta$ is the root-mean-square
surface roughness, and has no $k_{y}$-dependence. However, we do
not observe such anisotropy for $P_{n}^{\text{total}}$ in Figs.~\ref{fig:AllSpecularityParameters}(g)
to (i), indicating a disagreement with Soffer's formula. Furthermore,
in the long-wavelength limit, the $P_{n}^{\text{total}}$ for ZA phonons
does not converge to unity as suggested by the formula.

\subsection{Coherent vs. incoherent specularity parameters}

It is widely assumed~\citep{MNLuckyanova:Science12_Coherent,AMaznev:PRB15_Boundary,FShi:PRB17_Diffusely}
that coherent (incoherent) scattering is perfectly specular (diffuse),
i.e., $P_{n}^{\text{coh}}=1$ ($P_{n}^{\text{incoh}}=0$), although
there is no direct evidence for this relationship. Underlying this
assumption is the idea that the perfect interface is smooth although
at the atomistic level, lattice imperfections must occur because of
the crystallographic discontinuity. Given this assumption, it follows
from Eq.~(\ref{eq:CoherenceSpecularityRelation}) that coherence
is equivalent to specularity ($\mathcal{C}_{n}=P_{n}^{\text{total}}$).
We exploit our ability to distinguish coherent from incoherent scattering
to analyze how specularity actually depends on coherence, by comparing
the $P_{n}^{\text{coh}}$ and $P_{n}^{\text{incoh}}$ distributions
in Figs.~\ref{fig:AllSpecularityParameters}(j) to (o). The corresponding
$P_{n}^{\text{coh}}$ and $P_{n}^{\text{total}}$ distributions generally
have similar $\boldsymbol{k}_{n}$-dependence, with $P_{n}^{\text{coh}}>P_{n}^{\text{total}}$
because incoherent scattering is strongly diffuse ($P_{n}^{\text{incoh}}\ll1$)
with no significant $\boldsymbol{k}_{n}$-dependence for ZA, TA and
LA phonons, as can be seen in Figs.~\ref{fig:AllSpecularityParameters}(m)
to (o), and Eq.~(\ref{eq:CoherenceSpecularityRelation}) implies
that $P_{n}^{\text{total}}<\max\{P_{n}^{\text{coh}},P_{n}^{\text{incoh}}\}$.
The near uniform small value of $P_{n}^{\text{incoh}}$ over the entire
BZ in Figs.~\ref{fig:AllSpecularityParameters}(m) to (o) also suggests
that the diffuse character of incoherent scattering is captured by
Eq.~(\ref{eq:IncoherentSpecularityEstimate}).

\begin{figure}
\begin{centering}
\includegraphics[width=8cm]{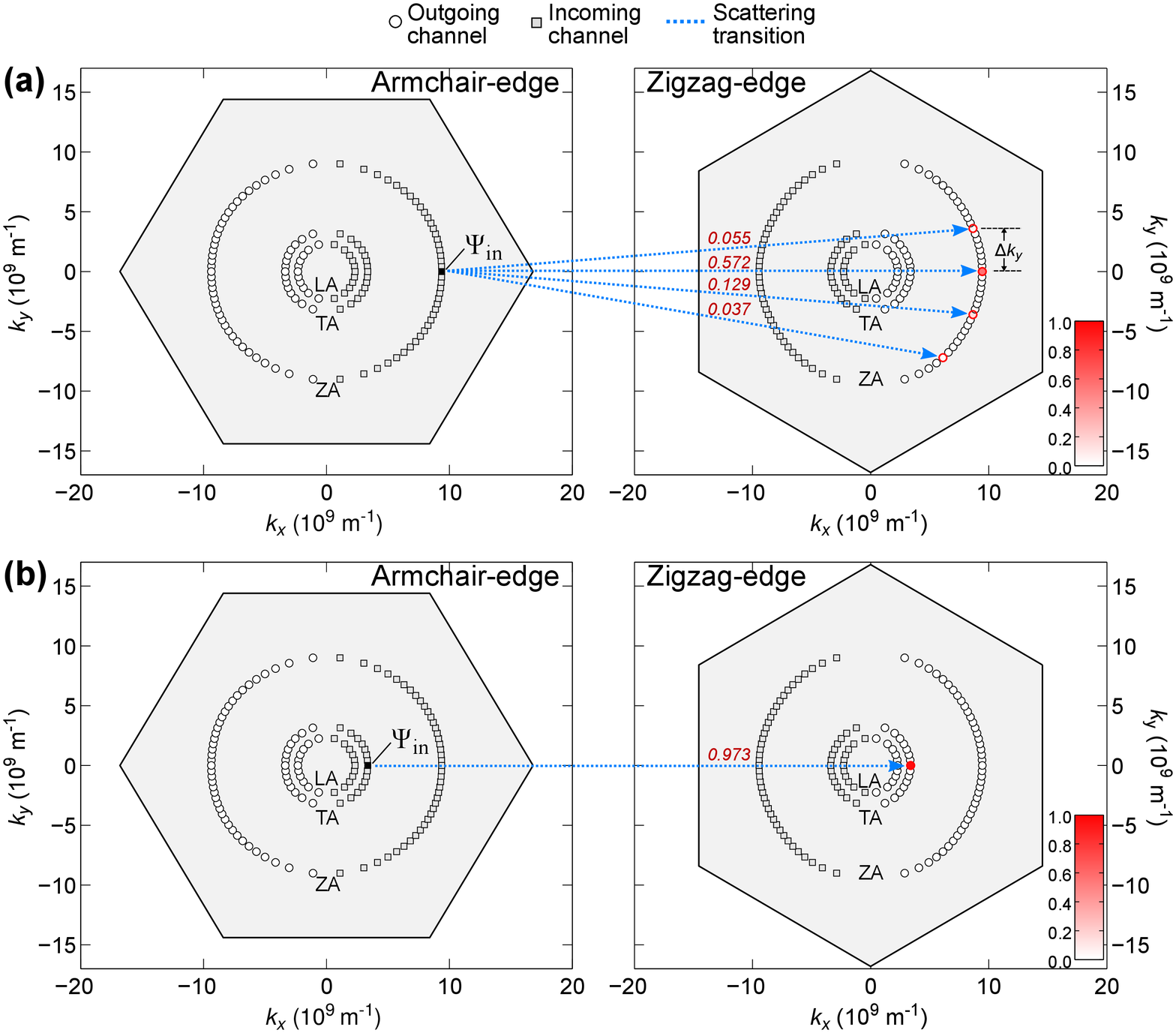}
\par\end{centering}
\caption{Main scattering transitions for an incoming \textbf{(a)} ZA and \textbf{(b)}
TA phonon, labeled $\Psi_{\text{in}}$, at normal incidence to the
grain boundary from the armchair-edge graphene on the left at $\omega=5\times10^{13}$
rad/s. The bulk LA, TA and ZA phonon channels on the armchair-edge
(left subpanel) and zigzag-edge (right subpanel) graphene side are
displayed within their respective first Brillouin zones. The color
scales indicate the transition probability from $\boldsymbol{W}_{\text{total}}(\omega)$
for the dominant outgoing channels, with the transitions indicated
by dotted lines and transition probabilities written in Italic font.}

\label{fig:ScatteringTransitions}
\end{figure}

Like in Fig.~\ref{fig:AllSpecularityParameters}(g), the $P_{n}^{\text{coh}}$
distribution for ZA phonons in Fig.~\ref{fig:AllSpecularityParameters}(j)
is significantly smaller than unity, indicating that even coherent
scattering is not fully specular for out-of-plane polarized phonons.
The $P_{n}^{\text{coh}}$ distribution for LA and TA phonons in Fig.~\ref{fig:AllSpecularityParameters}(k)
and (l) show that the coherent specularity diverges from unity as
we move away from the BZ center. To explain the reduced ZA phonon
specularity ($P_{n}^{\text{total}}$), we compare the main scattering
transitions for an incoming armchair-edge graphene (a) ZA and (b)
TA phonon at normal incidence ($k_{y}=0$) to the boundary at a single
frequency of $\omega=5\omega_{0}$ rad/s in Fig.~\ref{fig:ScatteringTransitions}.
The incoming ZA phonon is forward-scattered to several outgoing channels
while the incoming TA phonon is forward-scattered to a single outgoing
channel on the zigzag-edge side. The distinctive periodic arrangement
in the distribution of the main outgoing ZA phonon channels, separated
by an interval of $\Delta k_{y}$, is due to diffraction by the smooth
part of the boundary which has a periodicity equal to $W_{\text{GB}}$
the width of the constituent (4,4)|(7,0) GB such that $\Delta k_{y}=2\pi/W_{\text{GB}}$.
For a clear representation of diffraction by the `smooth' boundary
with the aforementioned transverse periodicity, we plot the equivalent
scattering transitions for the pure GB-II and GB-III boundaries in
Sec\@.~S6 of the Supplemental Material~\citep{ZYOng:SuppMater19_Structure}.
A similar effect has also been reported for molecular dynamics simulations
of symmetric graphene GB's~\citep{EHelgee:PRB15_Diffraction}. This
diffractive scattering is seen for other ZA phonon channels but none
of the in-plane LA and TA phonons.

\section{Summary}

We have formulated an $S$ matrix-based theory of boundary roughness
scattering to predict the mode-resolved coherence and specularity
parameters and applied it to the (32,32)|(56,0) graphene GB. The predicted
specularity parameters are shown to be consistent with those of Zhao
and Freund~\citep{HZhao:JAP09_Phonon}. We find that phonon scattering
is predominantly coherent for graphene GB's although contrary to expectations,
coherence and specularity are lowest for long-wavelength ZA phonons
because of diffractive scattering by the GB, while the opposite trend
is seen for LA and TA phonons. Our results also demonstrate that incoherent
scattering is much more diffuse than coherent scattering and that
coherence and specularity are not necessarily equivalent. Given its
generality, our method can be applied in a straightforward manner
to analyze phonon coherence and specularity in other atomistic boundary
models.
\begin{acknowledgments}
ZYO acknowledges financial support from a grant from the Science and
Engineering Research Council (Grant No. 152-70-00017) and the Agency
for Science, Technology, and Research (A{*}STAR), Singapore. GS acknowledges
support from EPSRC grant No.EP/J010863/2 and a grant from Tohoku University.
CJP is supported by the Royal Society through a Royal Society Wolfson
Research Merit award.
\end{acknowledgments}

\appendix

\bibliographystyle{apsrev4-1}
\bibliography{../manuscript_pnas/PaperReferences,SuppInfo}

\end{document}